\documentclass[journal=nalefd,manuscript=letter]{achemso}

\usepackage[version=3]{mhchem}

\mathchardef\mhyphen="2D

\usepackage{color}

\author{Hongliu Yang}
\affiliation[IfWW \& MBZ]
{Institute for Materials Science and Max Bergmann Center of Biomaterials (IfWW \& MBZ)}
\email{Hongliu.Yang@tu-dresden.de}
\author{Viktor Bezugly}
\affiliation[IfWW \& MBZ]
{Institute for Materials Science and Max Bergmann Center of Biomaterials (IfWW \& MBZ)}
\alsoaffiliation[cfAED]{Center for Advancing Electronics Dresden (cfAED)}
\author{Jens Kunstmann}
\affiliation[IfWW \& MBZ]
{Institute for Materials Science and Max Bergmann Center of Biomaterials (IfWW \& MBZ)}
\alsoaffiliation[TC]{Theoretical Chemistry (TC), Department of Chemistry and Food Chemistry, TU Dresden, 01062 Dresden, Germany}
\author{Arianna Filoramo}
\affiliation[CEA]
{DSM/IRAMIS/NIMBE/LICSEN, CEA de Saclay, 91191 Gif sur Yvette, France}
\author{Gianaurelio Cuniberti}
\affiliation[IfWW \& MBZ]
{Institute for Materials Science and Max Bergmann Center of Biomaterials (IfWW \& MBZ)}
\alsoaffiliation[DCCMS]{Dresden Center for Computational Materials Science (DCCMS), TU Dresden, 01062 Dresden, Germany}
\alsoaffiliation[cfAED]{Center for Advancing Electronics Dresden (cfAED)}

\title[An \textsf{achemso} demo]
  {Diameter-Selective Dispersion of Carbon Nanotubes via Polymers:
 A Competition between Adsorption and Bundling}

\abbreviations{ {\bf carbon nanotubes, diameter selectivity, polymer adsorption, binding energy, molecular dynamics, surface coverage}}

\begin{document}
\begin{abstract}
The mechanism of the selective dispersion of single-walled carbon nanotubes (CNTs) by polyfluorene polymers is studied in this paper. 
Using extensive molecular dynamics simulations, it is demonstrated that diameter selectivity is the result of a competition between bundling of CNTs and adsorption of polymers on CNT surfaces. 
The preference for certain diameters corresponds to local minima of the binding energy difference between these two processes. 
Such minima in the diameter dependence occur due to abrupt changes in the CNT's coverage with polymers
and their calculated positions are in quantitative agreement with preferred diameters, reported experimentally.
The presented approach defines a theoretical framework for the further understanding and improvement of dispersion/extraction processes.
\end{abstract}

Since their discovery, single-walled carbon nanotubes (CNTs) have attracted major research interest due to their extraordinary mechanical, chemical and electronic properties \cite{saito98}.
They are metallic or semiconducting depending on their chirality and as-synthesized material is normally a mixture of both types. 
For many applications, however, purified samples of only a certain type are in high demand. Purified semiconducting tubes are required, for instance, to achieve a large on/off ratio and high carrier mobility in thin-film field effect transistors \cite{chen05,lee11,bisri12,sangwan12,ding14,brady14} and
high power conversion efficiency for photovoltaics \cite{bindl10,holt10}. Moreover, for optoelectronic applications working in a specific wavelength range, the sorting of semiconducting CNTs according to diameter is of great importance.

In view of such demands, methods for the selective synthesis of CNTs of a certain electronic type or chiralites have been developed \cite{ding09,sanchez14}. A low-cost mass production of selected CNTs is yet to be achieved, however, and post-synthesis methods are often relied on \cite{hersam08}.
A promising post-synthesis selection method discovered recently is based on the
physisorption of polymers on the surface of CNTs, which has the advantage of leaving the electronic properties of CNT nearly unperturbed \cite{hersam08}. There is a relatively long history of using polymers to disperse CNTs in aqueous or organic solutions \cite{star01,oconnell01}. A recent finding is that, by using suitable 
polymers, CNTs can be selectively dispersed either for a specific diameter
range or for certain chiral angles \cite{hersam08}. Among those tested, the $\pi$-conjugated polymer
group of polyfluorene derivatives shows the ability to selectively disperse semi-conducting
CNTs \cite{nish07,chen07,gao11,ozawa11,tange12,gomulya13,mistry13,berton14,fukumaru14}. 
In particular, the di-octyl substituted polyfluorene (PFO) used with
toluene as solvent prefers to disperse small-diameter semi-conducting nanotubes with
chiral angles bigger than about 20 degrees \cite{nish07}. With longer side-chains, larger-diameter tubes can be dispersed but the chiral angle preference is gradually lost \cite{gomulya13,ding14}. More recently, copolymers of polyfluorene with anthracene or pyridine groups were found to selectively disperse large-diameter semiconducting nanotubes with high purity and high yields \cite{mistry13,berton14}. This fits well with the need to fabricate photoelectronic devices working in the infrared wavelength range \cite{avouris08}. 
Large-diameter nanotubes also benefit from a diminishing contact resistance and higher carrier mobilities.
Purified semiconducting CNTs have been used to fabricate high-performance field-effect transistors with high carrier mobilities and large on-off ratios \cite{lee11,bisri12,sangwan12,ding14,brady14}. 

Intensive experimental and numerical works have been undertaken to study the conformation of polymers adsorbed on the CNT surface. For DNAs and some bio-macromolecules \cite{zheng03}, which prefer to take a helical conformation even in the free state, a helically wrapped configuration on CNT was naturally expected. Studies on the adsorption conformation of linear conjugated polymers are less conclusive. For instance, poly(aryleneethynylene)s (PAE) were found to align linearly along the CNT when dispersed with toluene \cite{chen02}. The similar poly(p-phenyleneethynylene) polymer, poly[p-{2,5-bis(3-propoxysulfonicacidsodiumsalt)}phenylene]ethynylene, was found to form helically wrapped structures in aqueous dispersion \cite{kang09}. Imaged {\it via} SEM, it was shown that when dispersed in chloroform, poly(3-hexyl-thiophene) (P3HT) forms a helically wrapped structure on the surface of multi-walled CNTs \cite{giulianini09}. Recently, regioregular poly(3-alkylthiophene)s (rr-P3ATs) were used with toluene as solvent to 
enrich semi-conducting CNTs, and MD simulations showed that P3ATs take quasi-linear conformations, as adsorbed on CoMocat CNTs \cite{wang14}. In 2014 Shea \textit{et al.} reported the first experimental study on the adsorption configuration of PFOs on CNTs by using photoluminescence energy transfer and anisotropy measurements \cite{shea14}. Their data, however, are open to interpretation \cite{supp}.

In the past, efforts were also made to theoretically explain the selection mechanism. For DNAs, the intrinsic helical nature was believed to play a crucial role in their selective adsorption on CNTs \cite{zheng03}.
For aromatic polymers, Nish {\it et al.} \cite{nish07} found that PFOs on CNT surfaces form n-fold symmetric structures with their backbones aligned along the tube axis.
The magnitude of the binding energy between CNTs and polymers was shown to increase with the tube diameter, a trend that was later confirmed by several authors 
\cite{ozawa11,gomulya13,fukumaru14}. 
If the stability of adsorption complexes, as indicated by the binding energy, would determine the dispersibility of CNTs, the above results \cite{nish07,ozawa11,gomulya13,fukumaru14} would imply that large-diameter CNTs are more easily dispersed than small-diameter ones.
This, however, is in contradiction to experimental observations that PFO prefers to disperse small-diameter CNTs \cite{nish07,ozawa11,fukumaru14}. 
Furthermore, helically wrapped PFO structures on CNTs were used to explain the chirality preference of PFO \cite{gao11}. We will show below, however, that such helical structures are not dynamically stable.
Recently, a coarse-grained model was developed and used together with statistical mechanic arguments to explain the diameter preference of several pyridine-containing copolymers \cite{berton14}. But it is unclear how well the method can be transferred to other systems. 
Despite these advances, it is fair to say that a thorough understanding of the diameter and chirality selectivity of the polymer adsorption method is still lacking.

\section{Results and Discussion}
\begin{figure}[h!]
\centering
\includegraphics[scale=1.0]{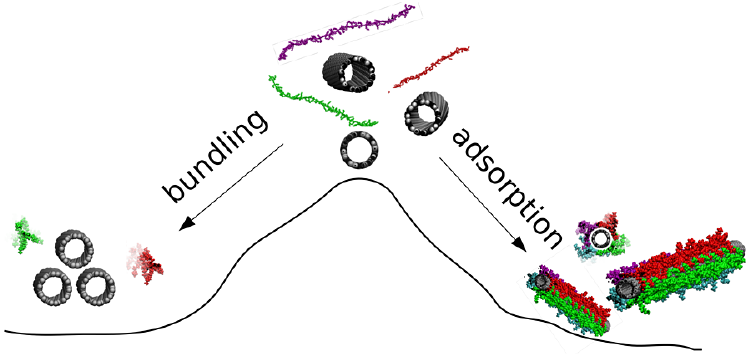}
\caption{The proposed mechanism of diameter-selective dispersion: a competition between bundling of carbon nanotubes (CNTs) and adsorption of polymers on the surface of CNTs. The initial state of the process, given by individual CNTs and individual polymers, is a transition state (on the top of a potential energy hill) created by sonication. 
}
\label{fgr:fig3}
\end{figure}

This article focuses
on understanding the diameter selectivity of the polymer adsorption method since the band gap and related electronic/optical properties of semiconducting CNTs are mainly determined by the diameter \cite{white93}. 
In particular, we propose that diameter selectivity results from a competition between the adsorption of polymers on the CNT surface and the bundling of individual CNTs (see Fig. \ref{fgr:fig3}). 
Our results on four relevant polymers are in excellent agreement with 
experimentally observed diameter preferences \cite{nish07,mistry13,berton14} and, thus, resolve a controversy on the nature of the mechanism that underlies the diameter selection process.
Despite the complexity of the competitive dispersion of CNTs, the success of our simple energetic model regarding diameter selectivity relies on its correct representation of some key factors including steric effects/coverage.

{\bf Simulations:}
To study the CNT-dispersion process, we performed classical molecular dynamics (MD) simulations using force fields.
Tip sonication treatment is known to generate high, local energy densities that break bundles into individual CNTs \cite{mason,huang12}. 
For the dilute polymer concentrations used in typical dispersion processes, the polymers exist as individual molecules \cite{justino11}. 
Therefore, isolated, individual CNTs and polymers were assumed as the initial configuration. Solvent molecules of toluene were usually not explicitly included here. 
We tested that their inclusion 
did not significantly change the results but mostly slowed down the adsorption dynamics.
Four representative types of polymers were considered in this study: the homopolymer of polyfluorene with side-chain length C8 (PFO) or C6 (PFH), and copolymers with anthracene group poly[(9,9-dihexylfluorenyl-2,7-diyl)-co-(9,10-anthracene)] (PFH-A),
or pyridine groups poly[9,9-didodecylfluorene-2,7-diyl-alt-pyridine-2,6-diyl] (PFD-Py). The chemical structures of these four polymers are presented in the Supporting Information.
Furthermore, 13 CNTs with diameters in the range from 0.8 to 1.4 nm
were considered. Such diameters are typically obtained in high-pressure CO conversion (HiPco) or pulsed laser vaporization (PLV) synthesis.
Further information on our simulation methods can be found in the Method section.

\begin{figure}[h!]
\centering
\includegraphics[scale=1.0]{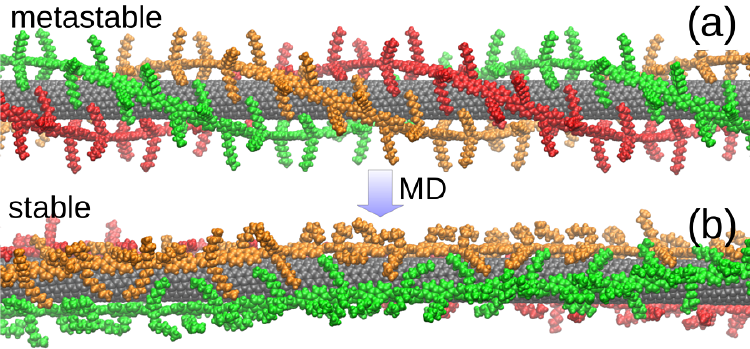}
\caption{The geometry of adsorption complexes: (a) A helically wrapped configuration is metastable, {\it i.e.},  a local minimum on the potential energy landscape. 
(b) A snapshot of the much more stable, linearly aligned configuration of PFO on a (8,6) CNT.
}
\label{fgr:fig1}
\end{figure}

{\bf Adsorption complexes:}
The geometries of adsorption complexes were obtained by MD simulations using many different initial configurations, temperatures, and CNT diameters. 
The simulations always lead to an almost linear alignment of PFO chains on the CNT surface, even after using initial conditions that promote the formation of helically wrapped structures.
Through geometry optimizations, we found that a multitude of such helically wrapped structures \cite{gao11}, with different pitches and surface coverages, are local minima on the potential energy landscape (see Fig.~\ref{fgr:fig1} (a) and S2) \cite{supp}. 
However, if they were subject to MD simulations
unwrapping proceeds gradually and after a sufficiently long run, a linearly aligned structure, as shown in Fig. \ref{fgr:fig1} (b), was always obtained. We conclude that helically wrapped adsorption complexes are metastable \cite{supp}. 

\begin{figure}[h!]
\centering
\includegraphics[scale=1.0]{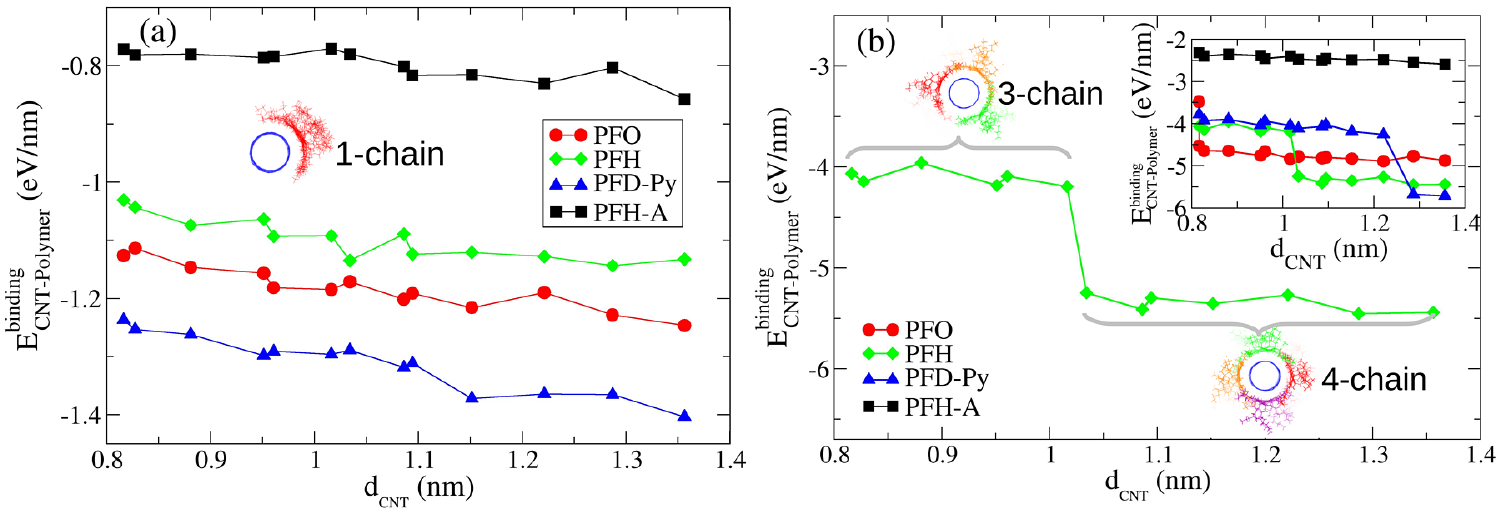}
\caption{The binding energy $E^{binding}_{CNT\mhyphen Polymer}$ (Eq.~\ref{eqn:ebin}) of adsorption complexes for 
(a) a single polymer chain and (b) the maximal coverage of the CNT surface. The magnitude of the binding energy increases with the nanotube diameter.
Therefore, the binding energy $E^{binding}_{CNT\mhyphen Polymer}$ alone cannot explain why polymers selectively disperse CNTs with specific diameters. 
Note the discontinuities of $E^{binding}_{CNT\mhyphen Polymer}$ in (b) that are due to abrupt changes in the surface coverage of the CNTs.
}
\label{fgr:fig2}
\end{figure}

{\bf Binding energy and stability of adsorption complexes:}
A standard measure used to characterize the stability of the adsorption complex is the binding energy. It is defined as the difference between the potential energy of an adsorption complex and the sum of its constituent molecules. For the adsorption of polymers on CNT, it reads
\begin{equation}
 E^{binding}_{CNT\mhyphen Polymer}=E_{CNT\mhyphen Polymer}-E_{CNT}-E_{Polymer}.\label{eqn:ebin}
\end{equation}
The binding energy for the adsorption of a single polymer chain on a CNT is shown in Fig.~\ref{fgr:fig2} (a). Note that the magnitude of the binding energy increases with the tube diameter in agreement with previous results \cite{nish07,ozawa11,gomulya13,fukumaru14}. 
This is caused by a better contact between polymers and CNT owing to the increasingly flatter surface of large-diameter CNTs \cite{supp}.
The side chain contributes a large part, about two-thirds, to the total binding energy of PFO. Consistently, the binding energy for PFH is smaller due to a shorter side-chain length \cite{lee11,gomulya13,wang14}.
The magnitude of the binding energy of 
PFH-A is smallest, which means that, for all the tested polymers with similar length, it is the easiest to remove from a CNT surface. This is in qualitative agreement with our recent experimental observation that PFH-A can be washed away from thin films deposited using dispersed CNTs (unpublished results). In contrast, PFO cannot be washed away in the same manner. 

{\bf Surface coverage of CNTs by polymers and binding energy of CNT-polymer complexes:}
To avoid the rebundling of CNTs after sonication, it is necessary to sufficiently cover the CNT surface with polymers.

We concentrate here on the situations where there is an excess of polymers and maximal coverage of the CNT surface is expected.
Binding energies for the maximal coverage of CNTs by polymers are shown in Fig.~\ref{fgr:fig2} (b). 
Note the discontinuities in the binding energy 
that are due to a sudden change in the number of polymers needed for maximal surface coverage \cite{nish07}. The positions of the discontinuities are different from those reported previously in the literature \cite{nish07,ozawa11,gomulya13,fukumaru14} because our MD simulations lead to different surface coverages 
than the geometry optimizations performed in those works \cite{supp}. 
These discontinuities have a direct relation to the diameter preference of polymers, as will be discussed below. 

{\bf Polymer-assisted dispersion as a competition between adsorption and bundling:}
The binding energy $E^{binding}_{CNT\mhyphen Polymer}$ alone cannot
explain the selectivity of the polymer adsorption method, because 
its magnitude simply increases with the diameter (see Fig. \ref{fgr:fig2}). 
This would imply that large-diameter CNTs are more easily dispersed than small-diameter ones. 
However, this is in clear contrast to the experimental observations discussed above \cite{nish07,mistry13}. The binding energy between polymer-wrapped CNTs could explain well the polymer-assisted dispersion of CNTs in certain solvents but not the selectivity on CNTs.
The key factor for understanding the selection mechanism
is competition between the bundling of CNTs on the one hand and the adsorption of polymers on the CNT surface on the other (see Fig. \ref{fgr:fig3}). 
This reasoning is based on the observation that CNT dispersions in toluene without polymers are not stable and the CNTs eventually rebundle. 
For this competition to take place, the initial state to be considered is a transition state consisting of individual polymers and individual CNTs. This transition state is experimentally realized by sonication, an integral work step of all selection methods.
Therefore, the selectivity of CNTs is determined by the \textit{difference} between the binding energy for CNT bundling and the binding energy for polymer adsorption, which reads 
\begin{equation}
\Delta E^{binding}= E^{binding}_{CNT\mhyphen Polymer} -  E^{binding}_{CNT\mhyphen CNT}. \label{eqn:ediff}
\end{equation}

\begin{figure}[h!]
\centering
\includegraphics[scale=1.0]{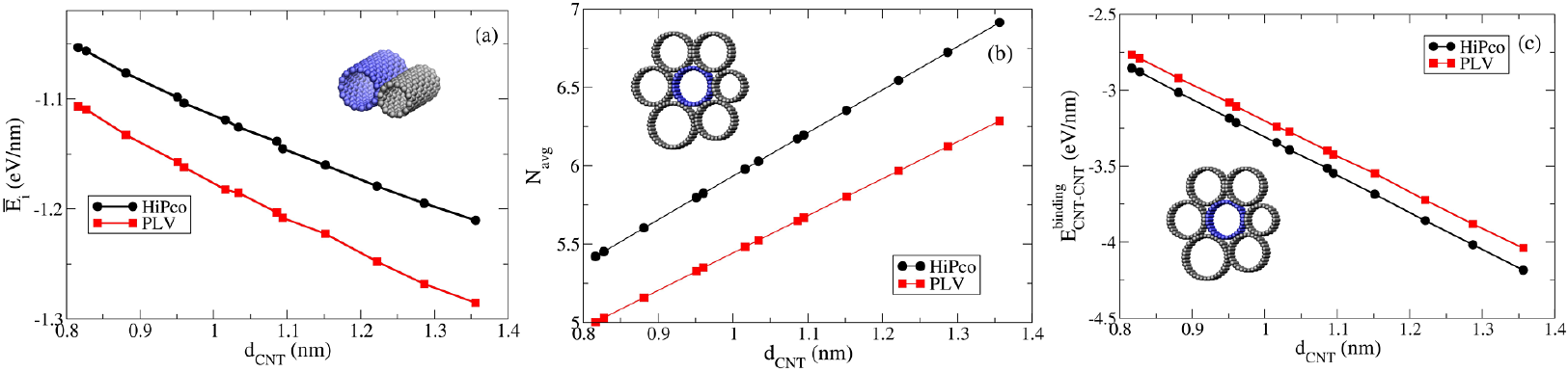}
\caption{The energetics of CNT bundling: (a) Weighted average $\bar E_i$ (Eq.~\ref{eqn:ecntpair}) of the pair binding energy of the interaction between a CNT of a given diameter and another CNT, arbitrarily selected from the sample. (b) Average number of neighbors $N_{avg}$ of a CNT with a given diameter in a bundle with mixed diameters (fractional numbers are a result of the non-uniform diameter distribution). 
(c) Binding energy $E^{binding}_{CNT\mhyphen CNT} = \bar E_i N_{avg}$ of CNT bundling. The variation of $E^{binding}_{CNT\mhyphen CNT}$ with the diameter follows the same trend as $E^{binding}_{CNT\mhyphen Polymer}$ in Fig.~\ref{fgr:fig2}, and only the competition between adsorption and bundling leads to selectivity. 
}
\label{fgr:fig5}
\end{figure}

{\bf Binding energy of CNT bundles:}
As-produced CNTs are normally a mixture of different diameters. This polydisperse nature makes a direct simulation of bundling computationally very expensive.
To overcome the difficulties, we first calculated the average pair binding energy $\bar E_i$ of the interaction between a CNT of a given diameter and another CNT, arbitrarily selected from a sample of mixed CNTs. 
It reads 
\begin{equation}
\bar E_i= \sum_j w_j  E_{ij}, \label{eqn:ecntpair}
\end{equation}
where $E_{ij}$ is the pair binding energy between CNT species $i$ and $j$, and $w_j$ is the population weight (abundance) of CNT species $j$ in the sample.
As shown in Fig. \ref{fgr:fig5} (a), the magnitude of $\bar E_i$ increases with the CNT diameter, due to the increase in the contact area between CNTs. 
Next, we estimate the average number of neighbors $N_{avg}$ of a CNT in bundles. A simple approach would be to ignore the polydispersity and assume that all CNTs have just six neighbors. But our method is to
consider the surface of a CNT of a given diameter to be covered by CNTs having the average diameter of the considered sample, {\it i.e.} HiPco or PLV.
Therefore, the average number of neighbors can be a non-integer. The estimates of $N_{avg}$ for two CNT samples are presented in Fig.~\ref{fgr:fig5} (b).
Finally, the binding energy for CNT bundling $E^{binding}_{CNT\mhyphen CNT}$ can be calculated as 
\begin{equation}
E^{binding}_{CNT\mhyphen CNT} = \bar E_i N_{avg}.
\label{eqn:ecntbundlg}
\end{equation}
As shown in Fig. \ref{fgr:fig5} (c) for both samples, the magnitude of the binding energy of CNT bundles increases with the CNT diameter.

\begin{figure}[h!]
\centering
\includegraphics[scale=1.0]{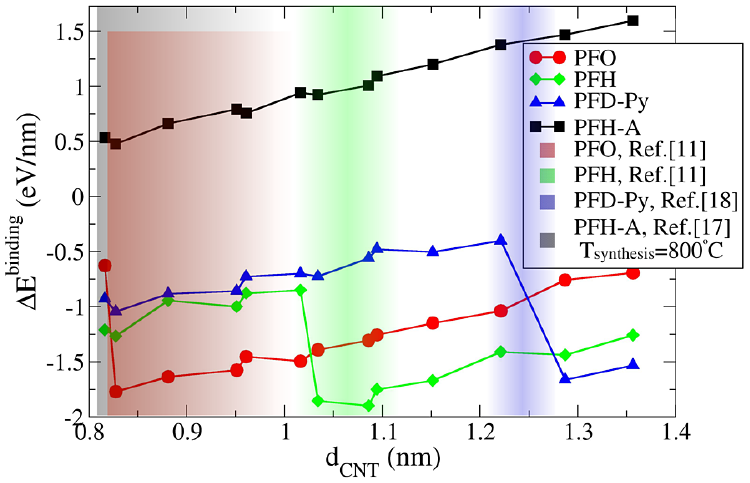}
\caption{Diameter selectivity as competition between CNT bundling and polymer adsorption. 
The diameter preferences of specific polymers for HiPco CNTs in our simulations are defined by the minima of the corresponding binding energy difference $\Delta E^{binding}$ (Eq.~\ref{eqn:ediff}). 
They are in excellent agreement with experimental results indicated by the shadowed regions
\cite{nish07,berton14,mistry13}. 
}
\label{fgr:fig6}
\end{figure}

{\bf Binding energy difference and diameter selectivity:}
As discussed already, for both CNT bundles and CNT-polymer complexes, the magnitude of the binding energy increases with the tube diameter. 
In the binding energy difference $\Delta E^{binding}$, the two trends 
nearly compensate for each other and only their competition leads to the preference for certain diameters, 
which are reflected in the location of the minima of $\Delta E^{binding}$ in Fig.~\ref{fgr:fig6}. 

Consider first the adsorption of PFO on HiPco CNTs in Fig \ref{fgr:fig6}. Except for the CNT with the smallest diameter, $\Delta E^{binding}$ increases with tube diameter. 
Since the number of polymers needed for the maximal surface coverage of the CNTs changes from three to four, $E^{binding}_{CNT\mhyphen Polymer}$ abruptly changes at 0.83 nm (see inset of Fig.~\ref{fgr:fig2} (b)), causing $\Delta E^{binding}$ to have a minimum at about the same diameter. 
The behavior of $\Delta E^{binding}$ explains 
(i) the preference of PFO to disperse HiPco CNTs in the diameter range 0.8--0.95 nm, a fact that has been repeatedly reported by different groups \cite{nish07,ozawa11,fukumaru14}, and
(ii) why CNTs with diameter smaller than 0.8 nm are not well-dispersed by PFO. These two insights explain the dominance of (8,6) CNTs (d = 0.95 nm) in HiPco CNT dispersions and the elimination of (6,5) CNTs (d = 0.75 nm) in CoMoCAT CNT dispersions \cite{nish07}. 

For PFH, with side-chains two carbon atoms shorter than PFO, more polymer chains are needed to cover the surface of a CNT. Therefore, the discontinuity in $\Delta E^{binding}$
is upshifted to 1.03 nm. This explains why,
for HiPco CNTs using PFH instead of PFO, the dominant CNTs in the dispersion become (8,7) (d = 1.02 nm) and (9,7) (d = 1.09 nm) (see Fig.1b of Nish {\it et al.} \cite{nish07}).

For the copolymer PFD-Py, the 
minimum of 
$\Delta E^{binding}$ is at about 1.25 nm. This agrees with recent experimental findings that, for HiPco CNTs, PFD-Py prefers to disperse CNTs with diameters of about 1.23 nm (see Fig.1n of Berton {\it et al.} \cite{berton14}).
 
The $\Delta E^{binding}$ of PFH-A increases continuously 
in the considered diameter range (0.8--1.4 nm)
and no minimum is discernible.
Mistry {\it et al.} performed a systematic study on the selectivity of PFH-A on CNTs synthesized {\it via} laser vaporization of graphite at different temperatures and found that
it always prefers to disperse CNTs with the smallest diameters in the sample \cite{mistry13}. The absence of a minimum in that range is again consistent with the experiments even though the simulations are based on HiPco CNTs. Further discussions on the selectivity of PFH-A can be found in the Supporting Information.

To summarize, for 
the polymers PFO, PFH and PFD-Py, the minima of the 
binding energy difference $\Delta E^{binding}$ 
match perfectly the experimentally reported diameters that are dominantly dispersed by those polymers.
This excellent agreement strongly suggests 
that the mechanism of diameter selectivity is a competition between CNT bundling and polymer adsorption.

It is interesting to note that the sign of $\Delta E^{binding}$ is negative for PFO, PFH and PFD-Py. This indicates a preference for the formation of CNT-polymer adsorption complexes over CNT-CNT bundling.
Therefore, for long sonication times, SWNTs of all diameters can be dispersed in principle. 
With increasing sonication time, the amount of dispersed CNTs will increase and the selectivity will gradually diminish.
Therefore, an optimal sonication time should be experimentally determined, providing a compromise between yield and purity.
Positive values of $\Delta E^{binding}$ for PFH-A mean that 
the rebundling should happen more frequently than adsorption,
which implies a potential lower yield of the dispersion process using PFH-A.

Note also that, due to the possibility of partial adsorption and other "imperfect" packing configurations the transition in Fig. 3(b) may turn out to be not so abrupt, see also the radial distribution functions shown in Fig.S6 and the corresponding discussions in the Supporting Information. Therefore, in experiments a range of diameters is often selected by a certain polymer.

One can also view the sonication-assisted dispersion process as a reversible reaction,
\begin{equation}
CNT@CNT + PFO \rightleftharpoons PFO@CNT . \label{eqn:ereac}
\end{equation} 
In this language, the initial configuration of isolated CNTs and polymers corresponds to a transition state, which is achieved with the aid of ultrasonication treatment \cite{mason,huang12}. 
The adsorption of polymers on the CNT surface and the bundling of CNTs are the 
rate-determining steps for forward and backward reactions, respectively. The two binding energies are the (negative) activation energies for reactions in the two directions. 
For this reversible reaction, the binding energy difference only estimates energetic contributions to
the reaction rates, neglecting 
entropic contributions, reaction orders and the concentrations of the reactants.

The focus of the current study is on the diameter selectivity of CNTs by aromatic polymers. For the chirality selection, the match/mismatch between the atomic structures of polymers and CNTs will be quite crucial. The popular implementations of van der Waals' interaction, as used here, were found unsuitable for the purpose and the anisotropic intermolecular potentials turn out to be a better alternative \cite{kc05}. For the sorting of CNTs with respect to electronic properties, {\it ab initio} quantum simulations with the electronic interactions being included would be more appropriate. All these issues deserve their own separate publications. The mentioned success of our simple energetic model implies that the key factors determining the diameter selectivity of (semi-conducting) CNTs were properly represented. The model can certainly be further improved by including: (i) entropy factors for a proper estimate of Gibbs free energy, which is of direct relevance to 
the reaction kinetics, and (ii) the effect of explicit solvent.
Calculations with explicit solvents and estimates of entropic contributions are provided in the Supporting Information.


In conclusion, we explain the diameter selectivity of polymer adsorption methods to be the result of a competition between the bundling of CNTs and the adsorption of polymers on the CNT surface. 
The preference of certain diameters corresponds to local minima of the binding energy difference between these two processes. Such minima occur due to abrupt changes in the CNT's coverage with polymers at certain diameters.
For all tested polymers including two homopolymers of polyfluorene with different side-chain lengths and two copolymers with anthracene or pyridine groups, our simulation results are in excellent agreement with the experimental findings regrading the diameter selectivity. Interestingly, even the influence of a fine-tuning of side-chain length on the selectivity was correctly captured in our method. Our insights resolve a long-standing controversy regarding the understanding of CNT selection schemes and 
are important for the further development of dispersion/extraction methods, {\it i.e.}, they enable MD simulations to be used for the screening of polymer candidates, tailoring of polymer structures, and obtaining further scientific insights. The proposed mechanism is general enough to be valid for other (sonication-aided) dispersion processes, for instance, the exfoliation of layered materials \cite{nicolosi13}, and the dispersion of CNTs by DNAs and mononucleotides \cite{zheng03,zheng09,ju08,johnson08}.

\section{Methods}

The adsorption of polymers on single-walled CNTs and the bundling of CNTs were studied with classical molecular dynamics (MD) simulations by using the CP2K \cite{cp2k} and Gromacs packages \cite{gromacs}. MD simulations were performed in NVT ensemble at $T=300K$ using the Nose-Hoover or Langevin thermostats.
The standard CHARMM force field parameters for the intra-molecular interaction \cite{charmm} were benchmarked against the density functional method (DFT) MD simulations with Grimme dispersion corrections DFT-D3 \cite{grimme10} and the BLYP exchange-correlation functional \cite{blyp} in CP2K and classical MD simulations with the MM3 force field \cite{mm3} using the Tinker package \cite{tinker}. The torsion angle parameter, describing amongst others the twist of the backbones of polymers, was modified to match the results of DFT first principles MD simulation. 
The inter-molecular interactions include an electrostatic part due to partial charges on atoms and a dispersion force part modeled by a Lennard-Jones potential as usual in standard CHARMM force field implementations \cite{charmm}.  

For the adsorption of polymers on the CNT surface, the polymer backbones were initially aligned parallel to the CNT axis. Multiple chains of polymers were arranged in an n-fold symmetric structure surrounding the tube. The initial distance between the backbone and the CNT surface was set to 1 to 1.5 nm depending on the CNT diameter and the number of polymers. For the binding energy calculation of CNT pairs, the two CNTs were placed in parallel with the initial distance between the surfaces of 0.6 nm. The time step for the integration of Newton's equation of motion was 1 fs. The duration of MD simulations ranged from 1 ns to 20 ns. For the calculation of thermodynamic averages, the equilibration time, ranging between 0.2 and 2 ns, was not considered.
To check the stability of self-constructed helical adsorption structures, geometry optimizations were performed using the CP2K package. The criteria of convergence are $3\times 10^{-3}$ $Bohr$ for the geometry change and $4.5\times 10^{-4}$ $Hartree/Bohr$ for the change in the force. 

It is known that the solvent plays an important role in the selective dispersion of CNTs by polymers \cite{huang08}. However, here we are interested in studying the effect of different polymers in combination with the same weakly polarized solvent, toluene.
Moreover, our tests showed that the explicit inclusion of solvent molecules of toluene in MD simulations did not change the structures of adsorption but significantly slowed down the dynamics of the adsorption process. Therefore, to enable MD simulations within reasonable times, solvent molecules were not explicitly included in our production runs. Our additional MD simulations with explicit solvents showed that, the adsorption configurations of the polymer backbones and sidechains, which are in contact with CNT surface, hardly change with the inclusion of solvents. Only the sidechains, which are not in contact with CNT surface, tend to point outwards to the solvent instead of folding back and aligning along CNT surface as in vacuum. The second-layer of polymers moves a bit away from the CNT surface. Therefore, with the inclusion of solvents the values of binding energies may change by some degree but the surface coverage of CNTs and the related positions of the abrupt changes in Fig.3(b) will be unaffected. Further details on the effect of explicit solvent can be found in the Supporting Information. 
Studies showed that a PFO octamer already has the same selectivity as a PFO polymer. Furthermore, the stability of the oligomer-CNT complex increases strongly with the chain length of the oligomer \cite{berton12}. 
To meet the capacity of the available computing resources in our simulation, we used 30-nm-long polymer 
chains, which consist of 32, 22 and 30 monomers of PFO/PFH, PFD-Py and PFH-A, respectively. 
CNT segments of length between 30 and 36 nm were used, their length varying with the chirality.  

To obtain the binding energy of a polymer-CNT complex, three MD simulations were
performed for the adsorption complex, the isolated CNT and the polymer, respectively.
The mean value of the potential energy was determined from the corresponding trajectories
and the binding energy was then calculated. This procedure is different from most cases in the literature where the binding energy was calculated from ($T=0$ K) single-point calculations of the optimized structures. 

It is worth pointing out that, for the adsorption of polymers on a CNT surface, the binding energy can be measured per unit length of a polymer chain, or per unit length of CNT covered completely by polymers. The former describes how hard it is to remove a polymer chain from the CNT surface while the latter is suitable for characterizing the competition for the adsorption on a CNT surface.

\begin{acknowledgement}
We acknowledge fruitful discussions with Gotthard Seifert. HLY thanks Jia Gao and Elton Carvalho for communications on the preparation of helically wrapped complexes. This work was partially funded by the European Union (ERDF) {\it via} the FP7 project CARbon nanoTube phOtONic devices on silicon (CARTOON) and is supported by Dresden Center for Computational Materials Science (DCCMS). We also acknowledge the support by the
German Research Foundation (DFG) within the Cluster of
Excellence ``Center for Advancing Electronics Dresden''
(cfAED).
We acknowledge the Center for Information Services and High Performance Computing (ZIH) at TU Dresden for computational resources.
\end{acknowledgement}

\begin{suppinfo}
Calibration of CHARMM force field parameters, stability of the helically wrapped configurations for PFO on CNTs, the determination of the surface coverage of CNT by PFO, variation of the binding energy of PFO-CNT complexes with CNT diameter, the population distributions of HiPco \cite{chen07} and PLV \cite{tange12} CNTs with respect to the diameter, the effect of solvent on the adsorption configurations, and the influence of entropic effect on diameter selectivity.
\end{suppinfo}


\begin{tocentry}

\centering
\includegraphics[scale=1.0]{scheme-competition2-b-resize.pdf}

\end{tocentry}

\end{document}